\documentclass[reprint,twocolumn,superscriptaddress,prb,showpacs]{revtex4-1}
\usepackage{graphicx}
\usepackage{dcolumn}
\usepackage{bm}
\usepackage{amsmath}
\usepackage{natbib}

\usepackage[T1]{fontenc}
\usepackage[letterpaper,textwidth=7in,top=.75in,bottom=.75in]{geometry}
\newcommand{\op}[1]{\hat{#1}}
\newcommand{\vekt}[1]{\mathbf{#1}}
\newcommand{\jav}{}

\begin{document}
\title{Optically Detected Magnetic Resonances of Nitrogen-Vacancy Ensembles in $^{13}$C Enriched Diamond}

 \author{A.~Jarmola}
  \email{jarmola@berkeley.edu}
    \affiliation{
     Department of Physics, University of California,
     Berkeley, California 94720-7300, USA
     }
     
 \author{Z.~Bodrog}
    \affiliation{
     Wigner Research Centre for Physics, Hungarian Academy of Sciences, PO Box 49, 1525 Budapest, Hungary
    }
    
\author{P.~Kehayias}
    \affiliation{
     Department of Physics, University of California,
     Berkeley, California 94720-7300, USA
     }

  \author{M.~Markham}
     \affiliation{
     Element Six, Fermi Avenue, Harwell, Oxfordshire, OX11 0QR, UK
    }

    \author{J.~Hall}
     \affiliation{
     Element Six, Fermi Avenue, Harwell, Oxfordshire, OX11 0QR, UK
    }
    
 \author{D.~J.~Twitchen}
     \affiliation{
     Element Six, Fermi Avenue, Harwell, Oxfordshire, OX11 0QR, UK
    }

\author{V.~M.~Acosta}
     \affiliation{
     Department of Physics and Astronomy, University of New Mexico, Albuquerque, New Mexico, 87106-4343, USA
     }

 \author{A.~Gali}
    \affiliation{
     Wigner Research Centre for Physics, Hungarian Academy of Sciences, PO Box 49, 1525 Budapest, Hungary
    }
    \affiliation{
     Department of Atomic Physics, Budapest University of Technology and Economics,
     Budafoki \'{u}t 8, H-1111, Budapest, Hungary
    }

 \author{D.~Budker}
 \email{budker@berkeley.edu}
    \affiliation{
     Department of Physics, University of California,
     Berkeley, California 94720-7300, USA
    }
    \affiliation{Helmholtz-Institut Mainz, Johannes Gutenberg Universität Mainz, 55128 Mainz,
Germany
    }

\date{\today}

\begin{abstract}

We present an experimental and theoretical study of the optically detected magnetic resonance signals for ensembles of negatively charged nitrogen-vacancy (NV) centers in $^{13}$C isotopically enriched single-crystal diamond. We observe four broad transition peaks with superimposed sharp features at zero magnetic field and study their dependence on applied magnetic field. A theoretical model that reproduces all qualitative features of these spectra is developed. Understanding the magnetic-resonance spectra of NV centers in isotopically enriched diamond is important for emerging applications in nuclear magnetic resonance.
\end{abstract}


\maketitle

\section{Introduction}

Recently, optical polarization of a $^{13}$C nuclear ensemble at room temperature was demonstrated with diamonds of natural isotopic abundance. \cite{FIS2013PRB, FIS2013PRL, KIN2015} The method is based on the transfer of optically induced electron spin polarization of negatively charged nitrogen-vacancy (NV) centers to the nuclei. The electron-nuclear spin transfer is enhanced when an external magnetic field is tuned to the NV excited-state \cite{JAC2009, SME2009} or ground-state \cite{WAN2013} level anticrossings. Synthesis of artificial diamonds affords the possibility of producing samples with modified isotopic abundance, with pure $^{12}$C diamond being one extreme, and pure $^{13}$C being the other. \cite{MIZ2009, NIZ2010B} The availability of $^{13}$C-enriched diamond with a high density of NV centers could enable realization of a fully nuclear-polarized solid at room temperature using optical cross-polarization technique. Importantly for many applications, such nuclear polarization techniques can be extended to nanodiamond, see for example, Ref.~[\onlinecite{CHE2015}]. The nuclear-spin environment in diamond samples can be characterized by optically detected magnetic resonance (ODMR) \cite{SIM2013} of the NV centers. As with previous observations \cite{NIZ2010B}, we observe that there are significant differences in ODMR spectra of natural-isotopic-abundance and $^{13}$C-enriched samples (Fig.~\ref{fig:ODMRzeroFiled}).

In this paper, we record ODMR spectra in samples with varying degree of $^{13}$C enrichment at different magnetic fields and develop a theoretical description of the observed spectral features. Understanding the ODMR spectra is an enabling step in developing efficient hyperpolarization techniques in isotopically modified diamond with implications for diamond based rotation sensors, \cite{LED2012, AJO2012} and applications in fundamental-physics research. \cite{LED2013}

\section{Experimental setup}

We used a standard confocal-microscopy setup to measure optically detected magnetic resonances in an ensemble of NV centers. The samples used in our experiments are single-crystal diamonds provided by Element Six and listed in Table \ref{tab:samples}. Sample 1 is an optical grade diamond grown using chemical vapor deposition (CVD) with a natural abundance (1.1\%) of $^{13}$C nuclear spins. Sample 2 is a high-pressure high-temperature (HPHT) grown diamond sample with 10\% enriched $^{13}$C concentration. Sample 3 is an electronic grade CVD diamond with a 30 $\mu$m thick layer of 99.9\% $^{13}$C isotopically enriched material. The NV centers in the $^{13}$C enriched layer were preferentially oriented along two of four possible crystallographic axes \cite{EDM2012, PHA2012PRB}. NV centers in samples 1 and 3 were formed during diamond growth. NV centers in sample 2 were created by irradiating it with 5 MeV electrons at a dose of 10$^{18}$ cm$^{-2}$ and subsequent annealing at 800~$^\circ$C for three hours. 532 nm laser light was focused on the diamond through a microscope objective with 0.6 numerical aperture. Fluorescence was collected through the same objective and detected with fiber-coupled single-photon counting module. A microwave (MW) field was applied using a 70 $\mu$m diameter copper wire placed on the diamond surface.

\begin{table}[ht]
\caption{\label{tab:samples} Sample characteristics. [$^{13}$C], [N], and [NV] indicate concentration of substitutional nitrogen, NV centers, and $^{13}$C, respectively. }
\centering
    \begin{ruledtabular}
    \begin{tabular}{c c c c c}
      \ Sample & Synthesis & [$^{13}$C] & [N] & [NV] \\
      \  No. &  & (\%) & (ppm) & (ppm)\\
      \colrule
      1 & CVD &  $1.1$ & $0.2-1$ & $0.005-0.01$ \\
      2 & HPHT & $10$ & $50-200$ & $5-20$ \\
      3 & CVD & $99.9$ & $0.2-1$ & $0.005-0.01$ \\
    \end{tabular}
    \end{ruledtabular}
\end{table}

\begin{figure}
\centering
    \includegraphics[width=1.0\columnwidth]{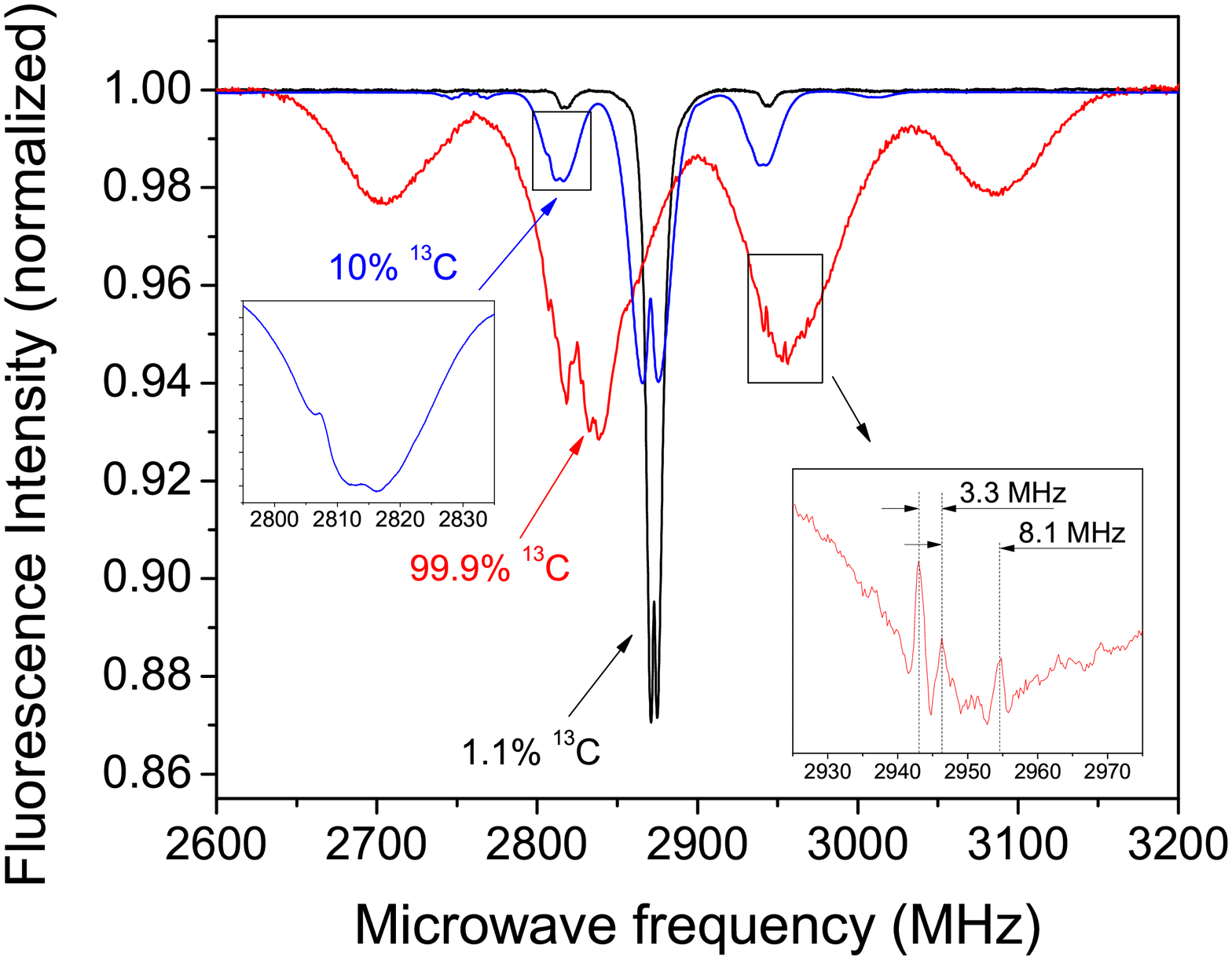}
    \caption{\label{fig:ODMRzeroFiled} Comparison of ODMR spectra for natural abundance of $^{13}$C (1.1\%) sample and enriched $^{13}$C isotopic abundance (10\%) and (99.9\%) samples at zero magnetic field. Narrow features on top of broad ones are observed for isotopically enriched samples (see insets).}
\end{figure}

Figure \ref{fig:ODMRzeroFiled} shows a comparison between NV ODMR spectra for the samples described above. The ODMR spectra for each sample are normalized individually to the fluorescence intensity obtained when the MW frequency is far detuned from the resonance. The spectrum for the natural-abundance sample shows a large central peak at the zero-field splitting between the $m_s=0$ and $m_s=\pm1$ magnetic sublevels of the ground state, with two asymmetrically offset (by $-$56 MHz and +70 MHz) smaller resonances corresponding to the presence of a single $^{13}$C nucleus in the ``first shell'' around the NV center (see Ref.~[\onlinecite{NIZ2010B}] and references therein). The central peak is split due to strain. For the 10\%-enriched sample, the overall structure of the spectrum is similar. However, the side resonances are larger (due to higher probability of finding a $^{13}$C nucleus in the first shell), all peaks are broader, and smaller additional peaks appear due to pairs of $^{13}$C nuclei in the first shell. The spectrum for the fully enriched sample is rather different. The central peak is not pronounced, while the spectrum consists of four irregularly shaped overlapping peaks with a substructure of significantly narrower features observed on the two central peaks.

\begin{figure}
\centering
    \includegraphics[width=1.0\columnwidth]{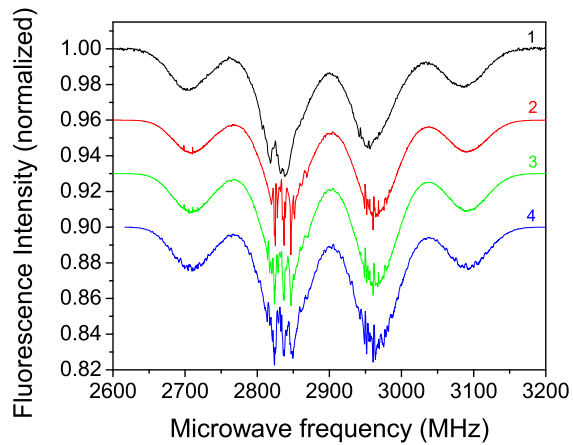}
    \caption{\label{fig:ODMRzeroFiled2} ODMR spectra of NV centers of an isotopically enriched (99.9\%) $^{13}$C sample at zero magnetic field. Curve 1 - experiment, curves (2-4) - simulations with various numbers of explicitly included proximal $^{13}$C nuclear spins: 2 - three nearest neighbors to the vacancy with hyperfine interaction A$_{\textrm{hfs}}$ = 130 MHz (first shell), 3 - first shell plus three $^{13}$C with  A$_{\textrm{hfs}}$ = 13.7 MHz, 4 - first shell plus six with A$_{\textrm{hfs}}$ = 13.7 MHz and two with A$_{\textrm{hfs}}$ = 12.8 MHz. The normalized spectra have been vertically offset for visual clarity.}
\end{figure}

\begin{figure}
\centering
    \includegraphics[width=1.0\columnwidth]{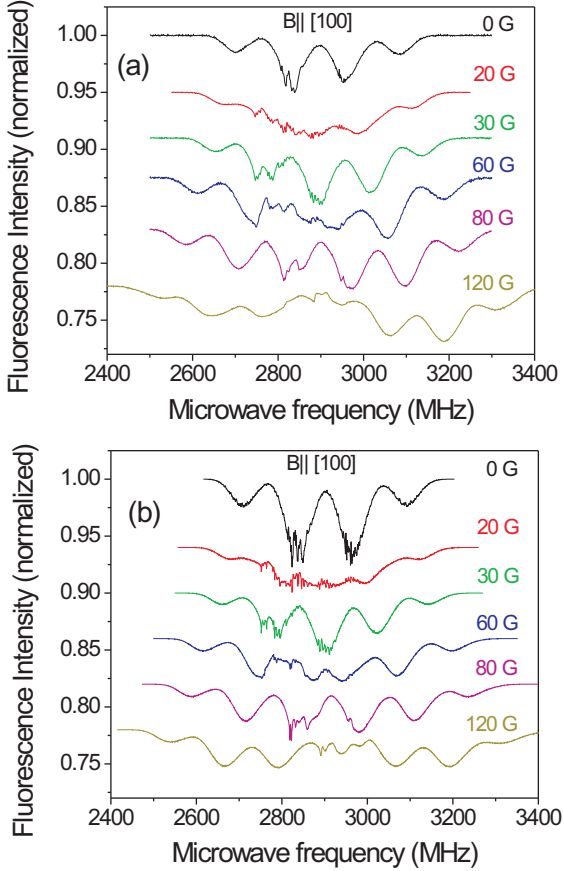}
    \caption{\label{fig:MagField} ODMR spectra of NV centers of an isotopically enriched (99.9\%) $^{13}$C sample for different values of magnetic field applied along [100] crystallographic direction. (a) experiment, (b) simulations. The normalized spectra have been vertically offset for visual clarity.}
\end{figure}

\section{Modeling of the spectra}
The basic structure of the NV ODMR spectrum for pure $^{13}$C diamond consisting of four main broad peaks can be understood as follows. The spins of the three first-shell $^{13}$C nuclei can add to a total spin of 1/2 or 3/2, leading to four possible projections ($-$3/2, $-$1/2, 1/2, 3/2) along the electron quantization axis. The four peak frequencies arise from the corresponding hyperfine shifts. The peaks are broadened by hyperfine interaction with further nuclei, which can be taken into account as a fluctuating magnetic background. To reproduce further details of the spectrum, e.g., the sharper features, a calculation is needed. To simulate the spectra produced by an ensemble of NV centers, we assumed that the values of the $E$ parameter describing the strain experienced by the individual NV centers follow a normal distribution with zero mean and a standard deviation of $\Delta E = 1$ MHz. The sample size was $10^{5}$, and we histogrammed the transition-energy values, which are the differences of the eigenvalues of the Hamiltonian.

To construct the Hamiltonian for the NV electronic spin triplet plus the
three neighboring carbon nuclei (one spin doublet for each $^{13}$C nucleus), we followed the method described by Nizovtsev \textit{et al}. \cite{NIZ2010} We used the anisotropic hyperfine interaction matrices for the nearest-neighbor carbons.

The Hamiltonian in the common space of states for the NV electrons and $^{13}$C nuclei contains elements like:
\begin{equation}
\label{eq:1}
\Psi = \psi \otimes \phi_{1} \otimes \dots
\otimes \phi_{n} \otimes \dots
\end{equation}
where $\psi$ denotes the NV electronic state and
$\phi_{n}$ is the $n^\text{th}$
${}^{13}\text{C}$ nuclear spin state. The Hamiltonian in this basis is
\begin{equation}
\label{eq:2}
\begin{split}
\op{H} &=
(D + \vekt{S} \cdot\vekt{B}) \otimes e_1 \otimes \dots \otimes e_n \otimes \dots
\\ &+
\sum_{ij} \alpha_{1,ij} S_i \otimes s_{1,j} \otimes \dots \otimes e_n \otimes \dots
\\ &+
\dots
\\ &+
\sum_{ij} \alpha_{n,ij} S_i \otimes e_1 \otimes \dots \otimes s_{n,j} \otimes \dots
\text{\jav{},}
\end{split}
\end{equation}
where $D$ is the zero-field-splitting (ZFS) tensor of the NV electronic triplet, $\vekt{S}$ is the
spin vector operator, $\vekt{B}$ is the external magnetic field (in frequency units, MHz), $e_n$ is the
$2\times 2$ unit matrix in the space of $^{13}$C nuclear spin $n$, $s_{n,j}$ is the $j^{th}$ component of nuclear spin $n$, and
$\alpha_{n,ij}$ is the matrix element of the hyperfine-interaction tensor
between NV electrons and the $n^\text{th}$ nucleus. We found that the [111] projection of the $\alpha_{n,ij}$ tensor is sufficient to construct the Hamiltonian in Eq.~(\ref{eq:2}) at high accuracy as the $D$ constant dominates the spin Hamiltonian and fixes the electronic spin direction.

Extending the above Hamiltonian to cover the second, third, etc.~neighbor nuclear spins is computationally
intensive, as the dimension of the Hamiltonian increases exponentially with the number of 
exact hyperfine interactions considered.  Given this, only the three first-neighbor hyperfine interactions were explicitly taken, and we accounted for the effect of farther nuclear spins by introducing a fluctuating
magnetic background. In order to calculate the magnitude of the fluctuating magnetic field caused by the $^{13}$C nuclear spins, we sum the effects of nuclear spins one by one of hyperfine constants larger than 8 MHz. We took the hyperfine interaction data from previous calculations.\cite{SME2009} These nuclei reside within 5~\AA\ from the vacancy of NV center. In this summation, the fluctuating fields of these nuclear spins are considered as independent random variables. The standard deviation of the magnetic field originating from each nucleus is given by the magnitude of its hyperfine splitting multiplied by the standard deviation of its freely fluctuating spin. 
The effect of more remote nuclear spins are averaged and summed up in spherical shells around the NV center as follows.

The component of the magnetic field coming from the $a^\text{th}$ shell is
\begin{equation}
\label{eq:3}
B_{a,x}(r_a) =
\frac{\sum_{i\in a} m_{i,x}}{r_a^3} -
3 \frac{\sum_{i\in a}(\vekt{m}_{i}\cdot\vekt{r}_i)r_{i,x}}{r_a^5}
\text{\jav{},}
\end{equation}
where $i$ is the label of each carbon atom in shell $a$, $r_a$ is the
radius of the shell, and $\vekt{r}_i$ is the position of atom $i$.
{\jav
Because the $\vekt{m}_i$ magnetic moments of individual ${}^{13}\text{C}$
nuclei are independent random variables of the same distribution, their
standard deviations add in quadrature:
{\jav
\begin{equation}
\label{eq:4}
\Delta B_{a,x} =
\sqrt{
\left(
\frac{\sqrt{N_a} \Delta m_x}{r_a^3}
\right)^2 + \left(
\sqrt{3} \frac{\sqrt{N_a} \Delta m_x}{r_a^3}
\right)^2
}
\text{.}
\end{equation}
}
Here $N_a$ is the number of carbon atoms in shell $a$, and we used
the fact tha{\jav{}t} $\Delta m_r = \Delta m_x$, i.e.\ the
{\jav standard deviation} of the
radial component of $\vekt{m}$ equals that of component $x$ due to spin fluctuation
isotropy. Finally, we also used a
{\jav `standard deviation'} of
$r_{a,x}/\sqrt{3}$ for the $x$ coordinate of atom $i$ ($\vekt{r}_i$ can be regarded as a pseudorandom variable, i.e. a deterministic set of points which is quite similar to a sample of uniformly distributed random points within the spherical shells they belong to). Combination of standard deviations within the parentheses takes into
account the fact that individual nuclear spins are independent of each other,
while the combination of the two parts in the parentheses, which are similar,
is a result of $m_r$ being the product of $m_x$ and a pseudorandom factor
independent of $m_x$.

If we take into account that the volume of a shell $a$ is
{\jav$4\pi r_a^2 dr$}, and the density of atoms in diamond is about $\varrho = 0.177$ per cubic
\AA, as well as that the {\jav standard deviation} of the $x$
component of an
isotropically distributed $s = 1/2$ spin is $1/2$, then the sum for all the shells can be approximated by an integral:
\begin{equation}
\label{eq:5}
\Delta B_x \approx
\xi
\sqrt{4\pi\varrho\int\limits_{r_0}^\infty \frac{1}{r^4}dr}
\text{,}
\end{equation}
where $\xi = g_I\mu_\text{N}g_S\mu_\text{B}\frac{\mu_0}{4\pi}
\approx 19.9\,\left[\text{MHz}\cdot\text{\AA}^3\right]$
($g_I$ is the gyromagnetic ratio of the ${}^{13}\text{C}$ nucleus,
$\mu_\text{N}$ is the nuclear magneton, $g_S$ is the electron spin
gyromagnetic ratio, $\mu_\text{B}$ is the Bohr magneton, $\mu_0$ is the
magnetic permeability of the vacuum).
Furthermore, if we set the inner radius of the first shell to $r_0 = 6$ \AA
(the last individually calculated nucleus is 5.05 \AA ~away), the final result is $\Delta B_{x} \approx$ 1.17 MHz.

The simulated and experimental ODMR spectra of 99.9\% $^{13}$C enriched CVD diamond are compared in Fig.~\ref{fig:ODMRzeroFiled2} with a varying number of proximal $^{13}$C nuclear spins that were explicitly taken into account (i.e. not in the magnetic background) in the simulation. In Trace 1 of Fig.~\ref{fig:ODMRzeroFiled2} the experimental curve shows some narrow features that are magnified in the inset of Fig.~\ref{fig:ODMRzeroFiled}. These narrow features are clearly visible in the most simple simulation (Trace 2 in Fig.~\ref{fig:ODMRzeroFiled2}) where only the first shell of $^{13}$C nuclear spins were explicitly taken in Eq. 2 and the rest were treated as a fluctuating background ($\Delta B_x \approx$ 24 MHz). The origin of the narrow features can be traced back in the simulation, and comes from a complex interplay of the random distribution of the $E$ constant (caused by strain in the sample) and the small magnetic field induced by the background nuclear spins. The  dips in the simulated curve of Trace 2 are more prominent than those in the experiment. By treating more $^{13}$C nuclear spins explicitly in the spin Hamiltonian (c.f., Traces 2, 3, and 4 in Fig.~\ref{fig:ODMRzeroFiled2} with an accordingly decreased fluctuating magnetic background $\Delta B_x$ of about 24 MHz, 17 MHz and 15 MHz, respectively) the narrow features are less and less sharpened and tend to converge to the experimental data. 
Our conclusion is that the experimental ODMR spectrum of NV center in $^{13}$C enriched diamond can be well understood by the relatively simple spin Hamiltonian given by Eq.~(\ref{eq:2}). 

The effect of a homogeneous external magnetic field can be straightforwardly introduced in this spin Hamiltonian. The experimental and calculated ODMR spectra with an external homogeneous magnetic field applied along the [100] direction are presented in Fig.~\ref{fig:MagField}. We choose this magnetic field direction because all four possible alignments of the NV centers with respect to the crystalline lattice are at the same angle to the magnetic field, and the spectra are generally the simplest. In these simulations we used our simplest model like Trace 2 in Fig.~\ref{fig:ODMRzeroFiled2}. In the presence of the external magnetic field, there is an ODMR resonance structure corresponding to each of the $m_s= 0 \rightarrow m_s =  \pm1$ transitions. These structures overlap, at least partially, up to the highest fields we have studied here. The basic features of the experimental spectra in the entire magnetic field range are excellently reproduced including the narrow resonances. Again, the overly sharp dips in the simulated spectra compared to the experimental data would be smoothened by the explicit inclusion of the proximate $^{13}$C nuclear spins of $>$8 MHz hyperfine couplings as shown in Fig.~\ref{fig:ODMRzeroFiled2}. This result further confirms the validity of our spin Hamiltonian in the description of NV center in $^{13}$C enriched diamond samples. We note that similar narrow resonances at higher magnetic fields have been recently studied in Ref.~[\onlinecite{PAR2015}].
We have also studied the dependence of the spectra on the parameter $\Delta E $. The modeling was done for 99.9\% $^{13}$C enriched sample in the range from 0.3 MHz to 3 MHz at zero magnetic field and at 30 G. The spectra remain largely the same including the width and position of the features. Bigger differences were observed for 10\% $^{13}$C  sample.

Conventional nuclear magnetic resonance (NMR) employs a strong (many tesla) magnetic field to thermally polarize nuclei at room temperature. Alternatively, at the NV  excited-state level anticrossing (near 500 G), NV centers can transfer optically pumped electronic polarization to nearby nuclei. \cite{JAC2009, SME2009} The eventual goal (and one motivation of our study) is to polarize all $^{13}$C in a diamond using NVs, which can be useful for NMR, quantum information, and sensing applications. \cite{FAL2015} In particular, $^{13}$C hyperpolarization would be especially useful if achievable in isotopically enriched diamond samples, such as those described in this work.

After measuring the ODMR spectra of samples 2 and 3 while stepping the magnetic field applied along [111] crystallographic direction in 3 G increments up to 575 G, we searched for evidence of $^{13}$C polarization with three methods:

\begin{itemize}
\item If the first-shell $^{13}$C nuclei are polarized, their corresponding ODMR peak depths will change to reflect a polarized population distribution.
\item If the non-first-shell $^{13}$C nuclei are polarized, this may lead to reduced ODMR linewidths, as the non-first-shell $^{13}$C magnetic fields are less random.
  \item If the bulk $^{13}$C nuclei are polarized, they will generate an additional magnetic field (a magnetization in the diamond). The NV transition frequencies will react to this additional field, and bulk polarization may appear as a nonlinearity in the $m_s =0 \leftrightarrow -1$ transition frequencies near 500~G.
\end{itemize}

Analyzing these parameters, we did not detect polarization in $^{13}$C enriched samples 2 and 3 (we note that hyperpolarization for 10\% enriched diamond was reported in Ref.~[\onlinecite{ALV2015}] which employed a different technique). This may be because transverse magnetic fields spoil the NV to $^{13}$C polarization transfer efficiency. The transverse fields are larger in enriched samples compared to a natural abundance ($\sim$100 MHz of magnetic inhomogeneous broadening in sample 3 compared to $\sim$1 MHz in sample 1).

Furthermore, it was recently shown \cite{IVA2015} that the polarization of $^{13}$C is sensitive to the relatively short coherence time of the electron spin in the excited state where the transfer of electron polarization to $^{13}$C nuclei is enhanced near the excited-state level anticrossing (ESLAC). The electron coherence time is reduced at increased $^{13}$C densities. In the present work, we have not attempted hyperpolarization near the ground-state level anticrossing (GSLAC). However, the analysis in Ref.~[\onlinecite{IVA2015}] suggests potential difficulties. Near the GSLAC one has to consider that the hyperfine tensors of $^{13}$C nuclei at different lattice sites generally induce different spin flip-flop processes at a given external magnetic field and each $^{13}$C nucleus requires different optimal conditions to effectively spin-polarize it.  

In conclusion, in this work, we observed ODMR spectra of an ensemble of NV centers in a natural-abundance, a $10\%$-enriched, and a pure $^{13}$C diamond. The pure $^{13}$C diamond displays at near-zero bias field, four broad transition peaks with superimposed sharp features. Building on the earlier work of Nizovtsev \textit{et al}., \cite{NIZ2010B} we develop a theoretical model that reproduces the qualitative features of these spectra, as well as the behavior at higher magnetic fields, where a variable number of peaks is observed. These results will be useful for future research into bulk nuclear hyperpolarization, where the ``holy grail'' is creation of macroscopic diamond samples with all nuclei polarized. Such samples are of interest in nuclear magnetic resonance spectroscopy and imaging, rotation sensing \cite{LED2012, AJO2012}, and fundamental-physics experiments.

The authors are grateful to Jeronimo Maze and Ran Fischer for useful discussions. This work was supported by AFOSR and the DARPA QuASAR program, and by DFG through the DIP program (FO 703/2-1). Z. B. and A. G. acknowledge support from the "Lendület" program of the Hungarian Academy of Sciences and and EU FP7 Grant No. 611143 (DIADEMS).

\end{document}